\documentclass{article}
\usepackage{arxiv}
\usepackage[utf8]{inputenc}
\usepackage[T1]{fontenc}
\usepackage{amsmath,amssymb,booktabs,graphicx,microtype,natbib,url}
\usepackage[hidelinks]{hyperref}
\graphicspath{{figures/}}

\title{PANDA-D: Interface-free determination of two dynamic contact angles from molecular-density and center-of-mass profiles}
\author{
\href{https://orcid.org/0009-0001-2142-3681}{Georgii Konstantinovich Vandyshev}\quad
\href{https://orcid.org/0009-0008-9460-8266}{Grigorii Stanislavovich Gorkaev}\\
\href{https://orcid.org/0000-0002-8286-0447}{Nikolay Dmitrievich Kondratyuk}\quad
\href{https://orcid.org/0000-0002-8379-2132}{Kirill Mironovich Gerke}\\
\href{https://orcid.org/0000-0002-5328-2860}{Ilia Vladimirovich Kopanichuk}\\[0.6ex]
\small Moscow Institute of Physics and Technology (National Research University), Dolgoprudny, Russia\\
\small Correspondence: \texttt{kopan239@gmail.com}
}
\date{October 2026}
\renewcommand{\shorttitle}{PANDA-D for dynamic contact angles}
\hypersetup{pdftitle={PANDA-D: Interface-free determination of two dynamic contact angles},
pdfauthor={Georgii K. Vandyshev, Grigorii S. Gorkaev, Nikolay D. Kondratyuk, Kirill M. Gerke, Ilia V. Kopanichuk},
pdfkeywords={dynamic contact angle, molecular dynamics, PANDA-D, nanopore, interface-free analysis}}

\begin{document}
\maketitle
\begin{abstract}
Contact angles are central descriptors of wetting and capillary behavior at solid--fluid interfaces. Dynamic contact angles in molecular simulations are commonly extracted from two-dimensional liquid--vapor contours and near-wall tangent fits. We introduce PANDA-D, an extension of Predicting Angle from Nanoscale Density Analysis (PANDA) that determines both contact angles of a driven liquid roll without extracting an interface. The density profile measures occupied layer width but does not identify which meniscus is advancing. The method resolves this ambiguity by jointly fitting density and the layerwise center-of-mass displacement relative to the roll. A circular-meniscus model maps these profiles to receding and advancing angles and an effective near-wall offset. On synthetic rolls with prescribed angles, the reconstruction error is $0.29^\circ$ root-mean-square. We then analyze force-driven Lennard--Jones rolls between crystalline walls. Joint fits reproduce both measured profiles, whereas density-only fits can match density while assigning incorrect asymmetry. The resulting angles agree with independently determined circle-fit angles to $0.84^\circ$ root-mean-square. Across four wall--fluid attractions, the receding angle decreases and the advancing angle increases with force, approximately linearly over the studied ranges. By eliminating contour thresholds and near-wall tangent extrapolation from the primary estimate, the two-profile approach enables reproducible, automated angle analysis across molecular trajectories.
\end{abstract}

\section{Introduction}
Wetting governs how a liquid occupies a surface or pore, and the contact angle is its most compact geometric descriptor. The equilibrium angle is tied to the interfacial free-energy balance, whereas a moving contact line generally exhibits distinct advancing and receding apparent angles~\citep{deGennes1985,Bonn2009,Snoeijer2013}. Those angles enter discussions of spreading, forced displacement, and capillary transport, but their value depends on the length scale and protocol at which a curved interface is observed. At the pore scale, even a small angular difference can alter the interpretation of multiphase invasion and wettability heterogeneity~\citep{Blunt2013,Gerke2021PoreScale,Ipatova2026Wettability}. A reproducible molecular-scale measurement is consequently useful both as an observable in its own right and as an input to coarse-grained models.

Dynamic wetting is difficult partly because a moving three-phase contact line couples processes across scales. The no-slip continuum construction has a local stress singularity~\citep{HuhScriven1971}; slip, contact-line kinetics, and viscous bending offer different descriptions of its regularization and the resulting apparent angle~\citep{BlakeHaynes1969,Voinov1976,Cox1986,ThompsonTroian1997,Qian2006,RenE2007}. Molecular-kinetic and hydrodynamic accounts need not identify the same local angle, and a measured angle should therefore be reported with its observation scale and wall convention~\citep{BlakeDeConinck2002,BlakeDeConinck2011,Keeler2022}. The present paper does not choose among these constitutive theories. Its narrower aim is to measure two angles consistently from particle trajectories before asking which dynamical law, if any, describes them.

Molecular dynamics (MD) can resolve the fluid, solid, and their fluctuating interface simultaneously. Early contact-line simulations established the importance of molecular slip and the sensitivity of dynamic angles to solid--liquid interactions~\citep{ThompsonRobbins1989,Bertrand2009}. Later simulations examined microscopic versus apparent angles and detailed molecular motion near the contact line~\citep{FernandezToledano2020,FernandezToledano2021}. In a confined bridge the two menisci can respond differently to forcing, while the liquid layers next to the walls need not resemble a sharp continuum interface~\citep{Wang2019,Lacis2022,Camargo2024}. Temperature-control algorithms may also change nonequilibrium observables, so a comparison of angles from different thermostat protocols requires care~\citep{BasconiShirts2013,RuizFranco2018,Heinz2026}.

The usual MD angle estimator first constructs a time-averaged two-dimensional density map, chooses an interface contour, fits a circle or another curve, and extrapolates its tangent to a specified wall plane. Such an estimator is physically intelligible and remains an important independent check. Yet the result can vary with the density threshold, the points excluded near the layered wall, the fit interval, and the location of the wall plane. These choices are especially consequential for nanodroplets and nanobridges, for which a few particle diameters are a substantial fraction of the meniscus height~\citep{Peng2012,Becker2014,Ravipati2018,Polovinkin2024,Camargo2024}. Thus agreement between two contour protocols is not automatically a unique ``true'' angle. Large simulation campaigns add a practical difficulty: robust extraction must be repeated without manually repairing each imperfect interface.

An alternative is to infer geometry from integrated one-dimensional observables. Our original Predicting Angle from Nanoscale Density Analysis (PANDA) recovered an equilibrium angle from the density profile of a liquid bridge in a slit, without selecting a two-dimensional liquid--vapor contour~\citep{Semenchuk2025PANDA}. A subsequent automated implementation improved reproducibility across configurations~\citep{Semenchuk2025PANDANN}. Related work extended interface-shape inference to liquid--liquid systems in cylindrical pores~\citep{Vandyshev2026}; simulations combining nanoscale wetting and roughness further motivate automated, clearly defined observables~\citep{Guskov2025Roughness}. These equilibrium approaches do not directly solve the driven case: a horizontal density profile integrated across the roll tells us its layer width but not whether the rear or front meniscus produced a given part of that width.

This is an identifiability problem rather than merely a noisy-fit problem. If the two menisci are exchanged, the one-dimensional density is unchanged. For small opposite angular shifts about equilibrium, their first-order effects on density cancel. A second observable must record the orientation of the occupied layer. We use its horizontal center of mass relative to the center of mass of the complete roll, retaining the favorable averaging of a profile while encoding left--right asymmetry. The pair of observables is linked to two circular apparent menisci by a geometric forward model; a common near-wall offset acknowledges that a sharp idealized boundary cannot coincide exactly with the molecular wall. The underlying concept is geometric, not a particular thermostat-dependent force balance.

We introduce this two-profile extension as PANDA-D and test it in three stages. First, synthetic rolls with known angles probe recovery over several equilibrium angles, filling fractions, and splittings. Second, equilibrium and force-driven Lennard--Jones rolls test whether the model follows actual molecular profiles and how its angles compare with independently fitted circles. Third, angle--force plots summarize the measured response at several solid--fluid attractions, without assigning a fitted friction or surface tension. The tests distinguish algorithmic recovery from MD-specific model discrepancy. PANDA-D's practical claim is an angle estimate that does not require contour extraction; close agreement with a circle fit is a consistency check, whereas proving universal superiority would require a separate matched benchmark with known truth.

\section{Methods}
\subsection{Molecular-dynamics system and sampling}
LAMMPS was used to simulate a Lennard--Jones liquid between crystalline solid surfaces~\citep{Thompson2022}. The liquid bridge spans the slit in the wall-normal $z$ direction, extends periodically along $y$, and moves along $x$ when driven. We call this translationally invariant bridge a \emph{liquid roll}. The model follows the truncated-and-shifted Lennard--Jones wetting framework used for nanoscale contact-angle studies~\citep{Becker2014}. Its pair potential is
\begin{equation}\label{eq:lj}
U_{ij}(r)=\begin{cases}
4\epsilon_{ij}\big[(\sigma_{ij}/r)^{12}-(\sigma_{ij}/r)^6\big]-U^{\rm LJ}_{ij}(r_c),&r<r_c,\\
0,&r\geq r_c
\end{cases},
\end{equation}
where $i$ and $j$ label fluid or solid species, $r$ is their separation, $\epsilon_{ij}$ and $\sigma_{ij}$ are the pair energy and length scales, $r_c=2.5\sigma_{ff}$ is the cutoff, and $U^{\rm LJ}_{ij}(r_c)$ is the unshifted potential at that cutoff. Reduced units use $\sigma_{ff}=\epsilon_{ff}=m_f=1$, with $m_f$ the fluid-particle mass. All $\sigma_{ij}$ are unity; $\epsilon_{ss}=100\epsilon_{ff}$, whereas the varied wall--fluid attraction is $\epsilon_{sf}^{*}=\epsilon_{sf}/\epsilon_{ff}$.

The walls are fcc (100) surfaces with lattice constant approximately $1.552\sigma_{ff}$ and fixed atoms. The lateral box dimensions are $L_x=72.944\sigma_{ff}$ and $L_y=18.624\sigma_{ff}$; the physical slit separation is $H=31.816\sigma_{ff}$. Periodic boundaries apply in all three directions, and the wall slab plus its periodic image form the slit. There are $N_f=14157$ fluid particles. A static calculation at each interaction strength supplies a separate equilibrium reference. The dynamic series analyzed below uses $\epsilon_{sf}^{*}=0.25$, $0.28$, $0.35$, and $0.42$.

\begin{figure}[htbp]
\centering\includegraphics[width=\textwidth]{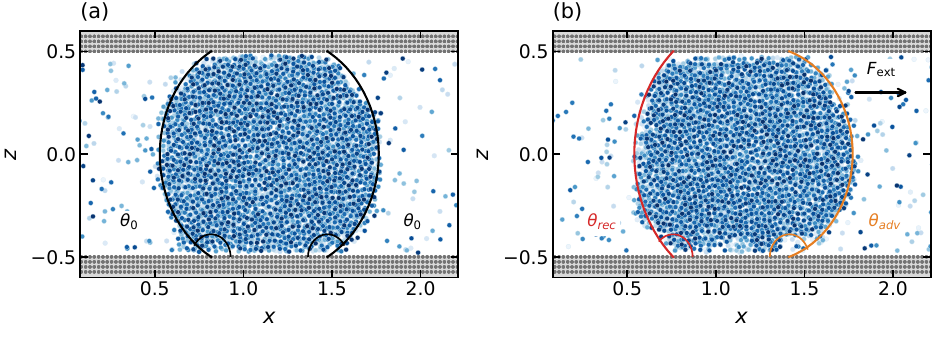}
\caption{Molecular configurations of the slit-spanning liquid roll at $\epsilon_{sf}^{*}=0.35$: (a) an equilibrium configuration and (b) a force-driven configuration. The plotted coordinates $x$ and $z$ are normalized by the pore height $H$, with $z=0$ at the pore center. Blue particles are fluid and gray particles are the fixed walls. In (b), the receding and advancing meniscus arcs and angle markers are red and orange, respectively. The imposed force points along $+x$. These schematic overlays are not the contour used by PANDA-D.}\label{fig:snapshots}
\end{figure}

The temperature is $T^{*}=k_{\rm B}T/\epsilon_{ff}=0.8$, and the time step is $0.002\tau$, with $\tau=\sigma_{ff}\sqrt{m_f/\epsilon_{ff}}$. The force-series plots use a Langevin thermostat~\citep{Brunger1984} acting on the transverse $y$ and $z$ velocity components, leaving the driven $x$ component outside the thermostat's temperature calculation. A spatially uniform force $f$ is applied in $+x$ to every fluid particle. Each trajectory comprises $70\times10^6$ steps, including $10\times10^6$ relaxation and $60\times10^6$ production steps. The angle analysis uses the late interval after step $40\times10^6$, divided into six temporal blocks. Other thermostat branches in the campaign are not combined with the four Langevin-$yz$ series here: thermostatting can affect nonequilibrium observables~\citep{RuizFranco2018}. Figure~\ref{fig:snapshots} displays an equilibrium roll and a driven configuration.

\section{Results and discussion}
\subsection{PANDA-D profiles and geometric model}
The equilibrium reference for an ideal, homogeneous roll can be motivated by minimizing interfacial free energy at fixed liquid volume. The liquid--vapor area and the liquid- and vapor-covered wall areas all contribute; their relative surface energies determine the equilibrium contact angle, while the stationary free interface has approximately constant curvature~\citep{deGennes1985,Bonn2009,Semenchuk2025PANDA}. For an extended roll in a slit, circular arcs are therefore a transparent low-dimensional \emph{geometric approximation} to the two menisci. The simplification to liquid--vapor area alone would require a special wetting limit and is not invoked for our partially wetting states. Nor do we assume that a driven steady state minimizes the equilibrium free energy: the model maps measured shape information to apparent angles, without deriving their force dependence.

Before profile construction, each saved frame is recentered on the roll with the periodic streamwise coordinate unwrapped consistently. In the geometric analysis all coordinates and distances are normalized by the physical pore height $H$; the resulting coordinates are denoted directly by $x$, $y$, and $z$. Thus $z=0$ is the pore center, the walls are at $z=\pm 1/2$, and the normalized lateral box lengths are $l=L_x/H$ and $l_y=L_y/H$. For a horizontal bin centered at $z_k$ with normalized thickness $\Delta z$, the number density in these scaled coordinates is
\begin{equation}\label{eq:rho}
\rho_k=\frac{N_k}{l\,l_y\,\Delta z},
\end{equation}
where $N_k$ is the number of fluid particles in bin $k$, and $l\,l_y\,\Delta z$ is its dimensionless full-box volume. Thus $\rho_k$ is a number density in units of inverse normalized volume; its ratio to the similarly defined liquid-core density does not depend on the choice of length unit. The profile includes vapor contributions where fluid particles occur outside the compact liquid region; the fit emphasizes the populated interior bins.

The second observable is the layerwise displacement
\begin{equation}\label{eq:q}
q_k=\frac{1}{N_k}\sum_{i\in k}\big(x_i-x_{\rm COM}\big),
\end{equation}
where $x_i$ is the normalized, unwrapped coordinate of particle $i$ in bin $k$, and $x_{\rm COM}$ is the normalized center of mass of the complete fluid roll in the same coordinate convention. Because the fluid particles have equal mass, the arithmetic layer mean is the layer center of mass. The quantity $q_k$ is undefined, rather than zero, for an empty bin; such bins are omitted from its fit. In the plots and fitting objective we use $R_k=\rho_k/\rho_l$ and $Q_k=q_k$, where $\rho_l$ is the liquid-core density in the same scaled coordinates.

The same centered, normalized $z$ coordinate enters the meniscus model. Let $h=1/2-\delta$ be the normalized half-height of the effective fluid region, where $\delta$ represents the near-wall layer inaccessible to the idealized meniscus. Let $d$ be the normalized separation of the two meniscus intersections with either effective wall. For side $i\in\{1,2\}$, the normalized horizontal bulge of a circular meniscus relative to its wall intersection is
\begin{equation}\label{eq:bulge}
b_i(z)=-\frac{\cos\theta_i\,(h^2-z^2)}{\sqrt{h^2-\cos^2\theta_i\,z^2}+h\sin\theta_i},\qquad |z|\leq h,
\end{equation}
where $\theta_1$ and $\theta_2$ are respectively the receding and advancing contact angles measured through the liquid. The expression is a numerically stable circular-arc parameterization; $b_i(\pm h)=0$ and positive $b_i(0)$ denotes a meniscus protruding away from the contact line for obtuse angles. Both angles are evaluated at the same wall plane after the fitted $\delta$ conversion.

The local roll width and normalized density predicted by the compact-liquid model are
\begin{equation}\label{eq:width}
w(z)=d+b_1(z)+b_2(z),\qquad R(z)=\frac{w(z)}{l},
\end{equation}
where $w$ is the normalized left-to-right liquid width, while $R$ is the fraction of the $x$ span occupied by liquid at height $z$. Thus the density profile depends on the \emph{sum} of the two meniscus bulges and cannot change when $\theta_1$ and $\theta_2$ are exchanged.

For later use we define the bulge moments and projected roll area,
\begin{equation}\label{eq:moments}
U_i=\int_{-h}^{h}b_i(z)\,dz,\qquad
V_i=\int_{-h}^{h}b_i^2(z)\,dz,\qquad
A=2hd+U_1+U_2,
\end{equation}
where $U_i$ and $V_i$ are respectively the integrals of the normalized bulge and its square over the normalized pore height, and $A$ is the normalized area occupied by liquid in an $xz$ cross-section. They enter the global center of mass of the idealized roll.

These full-height moments can be evaluated without numerical quadrature. Setting $k_i=|\cos\theta_i|$ and $s_i=\operatorname{sgn}(\theta_i-\pi/2)$ gives, for $k_i>0$,
\begin{equation}\label{eq:moments_closed}
\begin{aligned}
U_i&=s_i h^2\,
\frac{\arcsin k_i-k_i\sqrt{1-k_i^2}}{k_i^2},\\
V_i&=2h^3\,
\frac{k_i-\sqrt{1-k_i^2}\,\arcsin k_i-k_i^3/3}{k_i^3},
\end{aligned}
\end{equation}
where $k_i$ is the magnitude of the cosine of the $i$th angle and $s_i$ selects positive bulge area for an obtuse angle and negative bulge area for an acute angle; trigonometric arguments are in radians. The flat-meniscus limit is $U_i=V_i=0$ at $\theta_i=\pi/2$. Near that limit, $U_i=(2/3)s_i h^2k_i+O(k_i^3)$ and $V_i=(4/15)h^3k_i^2+O(k_i^4)$ provide stable evaluations when the closed forms suffer cancellation.

The predicted center-of-mass displacement is
\begin{equation}\label{eq:qmodel}
Q(z)=\frac{b_2(z)-b_1(z)}{2}
-\frac{d(U_2-U_1)+V_2-V_1}{2A},
\end{equation}
where the first term is the normalized horizontal center of the local layer relative to the midpoint of its two wall intersections; the second subtracts the normalized center of mass of the entire roll. The differences $U_2-U_1$ and $V_2-V_1$ retain the orientation information that the sum in Eq.~\eqref{eq:width} loses. The model assumes approximately uniform core density and circular apparent menisci; molecular layering is absorbed only approximately through $\delta$ and bin selection.

The projected liquid filling fraction $\phi$ and an independently fitted equilibrium angle $\theta_0$ determine the effective contact-line separation through
\begin{equation}\label{eq:d}
d=\frac{\phi l-2U(\theta_0,h)}{2h},
\end{equation}
where $U(\theta_0,h)$ is the integral in Eq.~\eqref{eq:moments} for a meniscus at the equilibrium angle, and $\phi l$ is the normalized projected liquid area inferred from the filling fraction. This construction anchors the symmetric reference state; $d$ is not separately inferred from a force-balance equation. In applying the model, $\theta_0$ is obtained from a same-interaction, same-thermostat zero-force PANDA-$\alpha$ analysis.

The reason to add $Q$ is especially clear for infinitesimal antisymmetric splitting, $\theta_1=\theta_0-\varepsilon$ and $\theta_2=\theta_0+\varepsilon$. At $\varepsilon=0$,
\begin{equation}\label{eq:sensitivity}
\left.\frac{\partial R}{\partial\varepsilon}\right|_0=0,\qquad
\left.\frac{\partial Q}{\partial\varepsilon}\right|_0=b_{\theta}(z)-\frac{dU_{\theta}+V_{\theta}}{A_0},
\end{equation}
where $b_{\theta}$, $U_{\theta}$, and $V_{\theta}$ are angle derivatives evaluated at $\theta_0$, and $A_0$ is the symmetric roll area. The first derivative of density cancels because the two bulges change in opposite directions; the center-of-mass profile retains a generally nonzero first-order signal. This is an identifiability result, not a dynamical force law.

PANDA-D fits the two measured profiles with a combined diagonal weighted objective,
\begin{equation}\label{eq:objective}
\chi^2=\sum_{k\in\mathcal I_{\rho}}\frac{[R(z_k)-\bar R_k]^2}{s_{R,k}^2}
+\sum_{k\in\mathcal I_q}\frac{[Q(z_k)-\bar Q_k]^2}{s_{Q,k}^2},
\end{equation}
where $\mathcal I_{\rho}$ and $\mathcal I_q$ are the valid interior-bin sets for density and center-of-mass displacement, overbars denote time-averaged observations, and $s_{R,k}$ and $s_{Q,k}$ are temporal-block scatter estimates with numerical floors to prevent a nearly noiseless bin from dominating. The optimized parameters are the two angles and the effective near-wall offset, with a multistart bounded search. No prescribed minimum angular gap or force-theory regularizer is used in the MD comparisons. Six-block standard deviations shown as error bars describe temporal variability rather than confidence intervals for systematic accuracy.

\subsection{Synthetic validation and analysis pipeline}
To test recovery against known angles, fluid coordinates were sampled inside prescribed circular-roll geometries. The test combines $\theta_0=110^\circ$, $130^\circ$, and $150^\circ$; filling fractions $\phi=0.3$, $0.5$, and $0.7$; and half-splittings $\Delta\theta=0^\circ$, $2^\circ$, $5^\circ$, $10^\circ$, and $20^\circ$. Each synthetic geometry uses $\theta_{\rm rec}=\theta_0-\Delta\theta$ and $\theta_{\rm adv}=\theta_0+\Delta\theta$, yielding 45 geometries and 90 individual true angles. Synthetic positions use the same order of particle number as the MD roll. This test checks the inverse problem under its assumed geometry; it does not include molecular layering or a diffuse molecular interface.

The analysis pipeline is fixed across synthetic and MD cases: periodically recenter every frame; accumulate $\rho_k$ and $q_k$ in wall-normal bins; estimate a liquid-core density and filling fraction; form the dimensionless profiles; determine the zero-force reference angle independently; fit the two angles and $\delta$ by bounded multistart optimization of Eq.~\eqref{eq:objective}; and inspect residuals, populated-bin support, and temporal-block variability. In the synthetic test, the prescribed geometry supplies the true reference and makes any inversion error directly observable. In MD, an independent two-dimensional circle analysis is reserved for validation, not fed into the PANDA-D objective.

\begin{figure}[htbp]
\centering\includegraphics[width=\textwidth]{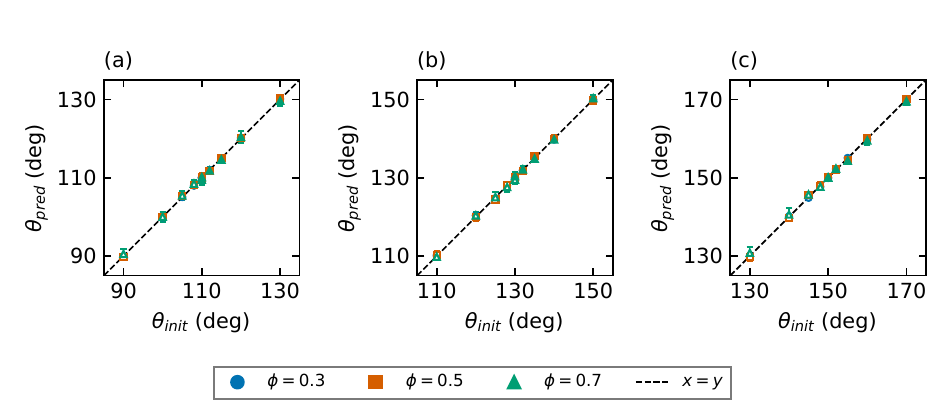}
\caption{Synthetic validation of PANDA-D at three equilibrium angles and three filling fractions. Prescribed receding and advancing angles are compared with those reconstructed from particle-derived density and center-of-mass profiles. The diagonal denotes exact recovery; the complete set contains 45 roll geometries and 90 individual angles.}\label{fig:synthetic}
\end{figure}

The synthetic benchmark separates algorithmic identifiability from MD-specific sources of bias. Across all 90 prescribed individual angles, the joint-fit root-mean-square error is
\begin{equation}\label{eq:rmse}
{\rm RMSE}_{\theta}=\left[\frac{1}{M}\sum_{j=1}^{M}
(\theta^{\rm fit}_j-\theta^{\rm set}_j)^2\right]^{1/2}=0.29^\circ,
\end{equation}
where $M=90$ is the number of receding and advancing angles combined, $\theta^{\rm set}_j$ is the prescribed input angle, and $\theta^{\rm fit}_j$ is its PANDA-D estimate. The errors for the $110^\circ$, $130^\circ$, and $150^\circ$ groups are $0.26^\circ$, $0.26^\circ$, and $0.34^\circ$, respectively. The points follow the equality line across all three filling fractions (Fig.~\ref{fig:synthetic}). This near-exact recovery is expected when the generated data obey the fitted circular geometry; it establishes that the two-profile inverse problem can be solved, not that real MD interfaces are circular to subdegree accuracy.

\subsection{Molecular-dynamics validation and comparison}
\begin{figure}[htbp]
\centering\includegraphics[width=0.54\textwidth]{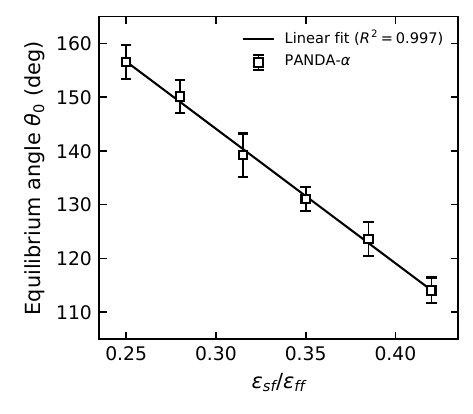}
\caption{Equilibrium PANDA-$\alpha$ contact angle for the well-sampled $\epsilon_{ff}=1$ series. Only the six long-run $\epsilon_{sf}^{*}=\epsilon_{sf}/\epsilon_{ff}$ points are shown; the two inserted interaction strengths, 0.315 and 0.385, each have duplicate trajectories. Error bars are temporal-block standard deviations. The black straight line is an unweighted linear least-squares fit of the means, not a physical theory.}\label{fig:equilibrium}
\end{figure}

The well-sampled $\epsilon_{ff}=1$ equilibrium series establishes a broad wettability range without mixing it with exploratory data at other fluid--fluid energies. At $\epsilon_{sf}^{*}=0.25$, $0.28$, $0.315$, $0.35$, $0.385$, and $0.42$, the zero-force PANDA-$\alpha$ angles are respectively $156.58^\circ$, $150.13^\circ$, $139.19^\circ$, $131.07^\circ$, $123.63^\circ$, and $114.03^\circ$ (Fig.~\ref{fig:equilibrium}). The four original points were analyzed over the last $30\times10^6$ steps of $70\times10^6$-step trajectories; each inserted $0.315$ and $0.385$ point was checked in two $45\times10^6$-step runs, again analyzing the last $30\times10^6$ steps. The plotted error bars are block standard deviations, not standard errors of the mean. The monotonic decrease of angle with stronger wall attraction agrees with the behavior expected of simple LJ wetting models~\citep{Becker2014,Semenchuk2025PANDA,Polovinkin2024}. An unweighted fit of the six means gives $\theta_0/\mathrm{deg}=219.21-250.30\epsilon_{sf}^{*}$ ($R^2=0.9974$). This linear relation only summarizes the sampled equilibrium range and is not a dynamic-wetting theory.

\begin{figure}[htbp]
\centering\includegraphics[width=0.92\textwidth]{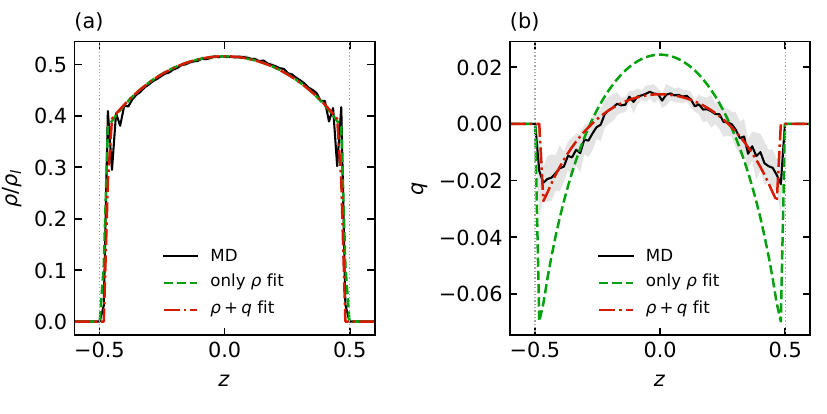}
\caption{Example MD profiles and fits: (a) normalized density and (b) normalized layerwise center-of-mass displacement. The plotted wall-normal coordinate $z$ is normalized by $H$. Black curves are measured time averages, gray bands represent temporal scatter, green dashed curves use density alone, and red dash-dotted curves use both profiles. The density-only solution fits panel (a) but misses the asymmetry measured in panel (b). Dotted vertical lines mark the wall positions; plotted zero extensions outside populated bins do not imply measured $q=0$ in empty bins.}\label{fig:profiles}
\end{figure}

The mathematical exchange symmetry of Eq.~\eqref{eq:width} has a direct numerical manifestation. In one representative driven case ($\epsilon_{sf}^{*}=0.35$, $f^{*}=2.5\times10^{-4}$), a density-only optimization and a joint optimization both track the observed normalized density to similar visual accuracy (Fig.~\ref{fig:profiles}a). Yet their predictions for $q$ differ markedly: the density-only solution exaggerates its central displacement and near-wall shape, while the joint fit follows the observed profile within much of the temporal scatter (Fig.~\ref{fig:profiles}b). Thus a good one-dimensional density fit is not sufficient evidence that the front and rear contact angles have been recovered. The second profile supplies orientation information without requiring a contour or local tangent. This methodological distinction extends the equilibrium PANDA approach rather than replacing its useful density constraint~\citep{Semenchuk2025PANDA,Semenchuk2025PANDANN}. Conventional contour approaches extract the missing orientation directly from a two-dimensional boundary~\citep{Peng2012,Camargo2024}; PANDA-D instead recovers it from an integrated particle observable.

For an independent geometric comparator, the time-averaged $xz$ density map is recentered and an interface is extracted at half the liquid-core density. Contour points within a normalized distance of approximately $0.094$ from either wall (three fluid-particle diameters) are excluded; separate circles with vertical centers fixed at $z=0$ are fitted to the two remaining contours. Their tangents are evaluated at the same wall planes, $z=\pm 1/2$, as the PANDA-D angles. This comparator does not enter the one-dimensional fit, but it is sensitive to contour, exclusion, and wall-position choices~\citep{Peng2012,Ravipati2018,Camargo2024}. Agreement with it tests consistency; the circle fit is not ground truth.

\begin{figure}[htbp]
\centering\includegraphics[width=0.94\textwidth]{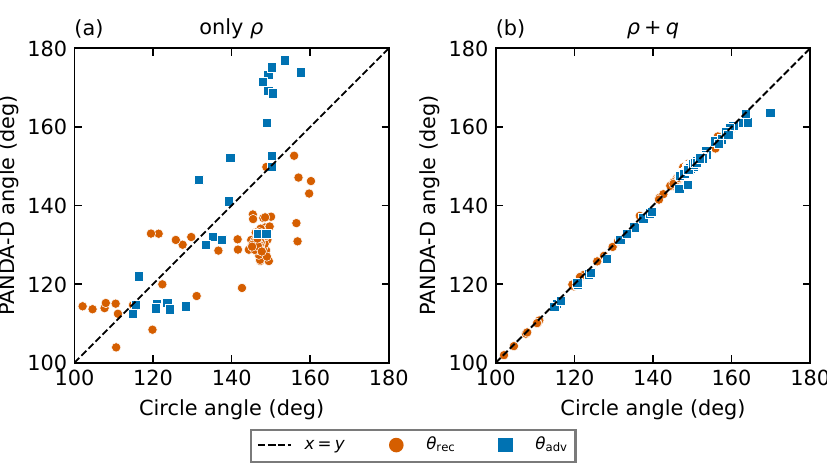}
\caption{PANDA-family estimates versus independent circle-fit contact angles. (a) Historical density-only control with an older equilibrium reference; (b) refreshed density-plus-center-of-mass PANDA-D estimate with the corrected same-mode reference. The dashed line is equality. Different accepted sample sizes and reference conventions make the panels illustrative rather than a controlled ablation.}\label{fig:comparison}
\end{figure}

Across the refreshed campaign, the PANDA-D joint-fit estimates lie close to independent circle-fit angles over a broad range of advancing and receding states (Fig.~\ref{fig:comparison}b). For the 167 quality-accepted individual angles in this comparison, the root-mean-square PANDA-D--circle difference is $0.84^\circ$. This number measures agreement between two methods on selected data, not absolute accuracy. Cases without a resolved circle contour or otherwise failing the comparison's quality rules are absent from this statistic.

The density-only control in Fig.~\ref{fig:comparison}a is a \emph{historical} analysis using an older equilibrium-angle reference, whereas the joint-fit panel uses the updated same-mode reference and wall conversion. The historical set contains 109 accepted individual angles and has a $14.80^\circ$ root-mean-square circle discrepancy. Because both the data selection and reference treatment differ, the two RMSE values must not be interpreted as a controlled effect size for adding $q$. The stronger evidence for the necessity of $q$ is the exact exchange symmetry of the density model, its vanishing first-order sensitivity in Eq.~\eqref{eq:sensitivity}, and the same-trajectory profile counterexample in Fig.~\ref{fig:profiles}. The practical advantage over circle fitting is procedural: PANDA-D's primary estimate does not depend on choosing a two-dimensional density threshold, contour segment, or extrapolated tangent. This does not prove that its angles are universally more accurate than circle fits. Direct superiority under realistic molecular noise requires an additional matched, known-truth benchmark.

\begin{figure}[htbp]
\centering\includegraphics[width=\textwidth]{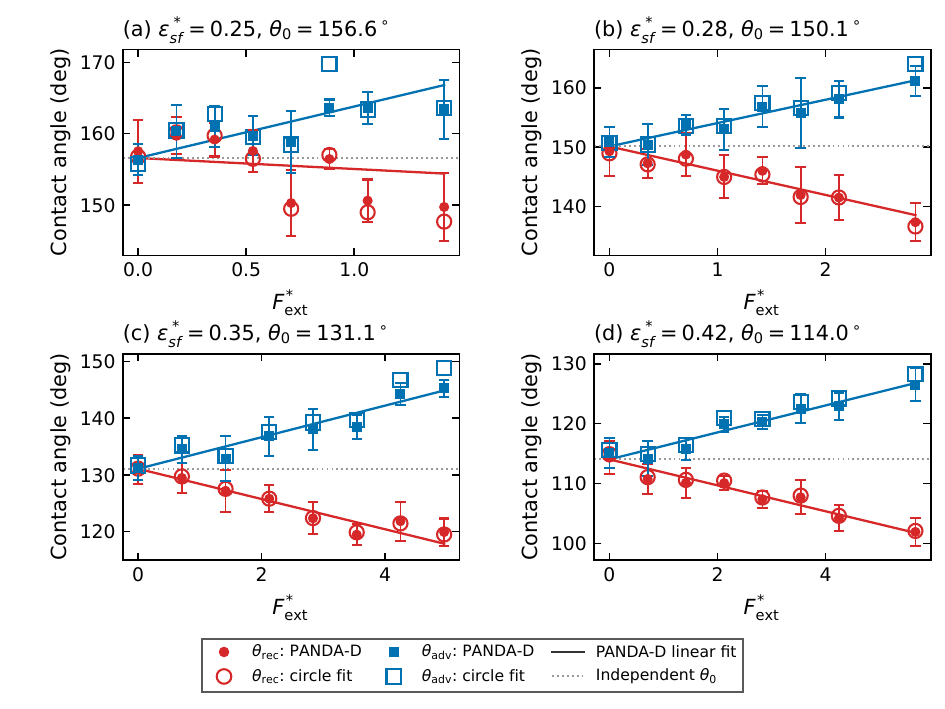}
\caption{PANDA-D receding (filled red circles) and advancing (filled blue squares) angles versus total reduced force for Langevin-$yz$ rolls at four wall--fluid interaction strengths. Open symbols show independent circle-fit estimates from the same trajectories; circle-fit uncertainty bars are not available in this comparison. PANDA-D error bars are six-block standard deviations. Colored lines are weighted empirical PANDA-D fits $\theta_s=\theta_0+k_sF_{\rm ext}^{*}$ with the intercept fixed to the independent zero-force PANDA-$\alpha$ result (gray dotted line). No force-balance theory is plotted.}\label{fig:response}
\end{figure}

Once both angles are identifiable, the four Langevin-$yz$ series show the expected opposite trends: the receding angle generally decreases, and the advancing angle increases, as the roll is driven more strongly (Fig.~\ref{fig:response}). A split between front and rear angles is familiar from moving-contact-line simulations, although its magnitude and detailed mechanism depend on the system~\citep{ThompsonRobbins1989,Bertrand2009,FernandezToledano2021,Wang2019}. Independent circle-fit estimates for the same trajectories are superimposed as open symbols; all 32 plotted states have both circle angles resolved. Their root-mean-square difference from PANDA-D is $1.24^\circ$ over the 64 angles in these four series. The empirical regressions use PANDA-D values only. For a compact description, each PANDA-D angle is fitted by a weighted linear line constrained to the independently determined zero-force angle,
\begin{equation}\label{eq:linear}
\theta_s(F_{\rm ext}^{*})=\theta_0+k_sF_{\rm ext}^{*},\qquad F_{\rm ext}^{*}=N_ff^{*},
\end{equation}
where $s\in\{\rm rec,adv\}$ labels the two interfaces, $\theta_0$ is the same-mode zero-force angle, $k_s$ is the empirical slope in degrees per reduced total force, $N_f$ is the number of forced fluid particles, and $f^{*}$ is the reduced force per particle. These lines are descriptive regressions, not theoretical predictions, and should not be extrapolated beyond the simulated force ranges.

At $\epsilon_{sf}^{*}=0.28$, the slopes are $k_{\rm rec}=-4.09$ and $k_{\rm adv}=+3.93$ degrees per force unit; at $0.35$ they are $-2.67$ and $+2.79$, and at $0.42$ they are $-2.17$ and $+2.25$. The weakly attractive $0.25$ series is more scattered, particularly for the receding angle: its fitted slopes are $-1.54$ and $+7.22$. The wall-affinity dependence of these \emph{observed} slopes is relevant to nanoscale wetting dynamics, but slope differences alone cannot identify molecular contact-line friction or separate it from thermostat and interfacial-shape effects~\citep{Snoeijer2013,FernandezToledano2021,RuizFranco2018}. We therefore avoid assigning a constitutive interpretation in this method paper.

PANDA-D gains information by compressing three-dimensional particle coordinates into two complementary one-dimensional profiles. This makes it attractive for automated trajectory analysis, but the circular-meniscus and uniform-core assumptions remain explicit. A noncircular meniscus, strong wall layering, substantial vapor populations, varying local density, or a contact line too close to $180^\circ$ can bias the geometric interpretation. Likewise, the apparent contact angle depends on the chosen physical wall plane even when the interface itself is not extracted. The fitted near-wall offset partly addresses this issue but cannot eliminate it. Block scatter omits model discrepancy, contour-systematic uncertainty, and uncertainty in the equilibrium reference. The method should therefore report profile residuals, valid-bin support, convergence information, and an independent geometric comparison when one is available. The selected cases demonstrate feasibility for a simple roll, not universal performance for arbitrary contact-line topologies. A next step is to transfer the analysis to GROMACS for chemically realistic fluids and substrates and validate it for more complex meniscus shapes.

\section{Conclusions}
PANDA-D turns the asymmetric geometry of a moving liquid roll into a two-profile inverse problem. The density profile alone cannot distinguish the advancing from the receding meniscus, whereas the layerwise center-of-mass profile supplies the missing orientation information. Joint analysis therefore yields both dynamic contact angles without selecting a liquid--vapor contour for the primary estimate.

On synthetic configurations, PANDA-D recovered prescribed angles with a root-mean-square error of $0.29^\circ$. For force-driven Lennard--Jones rolls, the joint model reproduced both measured profiles, while density-only fits could match density but miss the observed asymmetry. PANDA-D angles agreed with independent circle fits to $0.84^\circ$ root-mean-square. Across four wall--fluid attractions, the receding angle decreased and the advancing angle increased with force, approximately linearly over the investigated ranges.

These results establish PANDA-D as a practical, reproducible method for extracting dynamic contact angles from molecular trajectories with less dependence on interface detection. Future work will test more complex meniscus shapes and transfer the analysis to GROMACS simulations of chemically realistic fluids and substrates.

\section*{Acknowledgments}
This work was supported by the Russian Science Foundation (project No. 26-73-10086). The computations were performed using the Soft Cluster and CCP resources of the Center for Computational Physics (MIPT).

\section*{Data and code availability}
The data that support the findings of this study are available from the corresponding author upon reasonable request

\bibliographystyle{unsrtnat}
\bibliography{references}
\end{document}